\documentclass[12pt,a4paper]{conference}
\usepackage{fancyhdr}
\usepackage{graphicx,amsmath,amssymb,cite}
\usepackage{multind}
\makeindex{author} 
\makeindex{subject}

\RequirePackage{xspace}
\def\kaos{{\sc Kaos}$/\!${\small A1}\xspace}

\newcommand{\beq}{\begin{equation}}
\newcommand{\eeq}[1]{\label{#1}\end{equation}}
\newcommand{\eeqn}{\end{equation}}

\newcommand{\beqa}{\begin{eqnarray}}
\newcommand{\eeqa}[1]{\label{#1}\end{eqnarray}}
\newcommand{\eeqan}{\end{eqnarray}}

\let\bar=\overbar

\newcommand{\Dslash}{\not{\hbox{\kern-4pt $D$}}}
\newcommand{\dslash}{\not{\hbox{\kern-2pt $\del$}}}

\newcommand{\msb}{{\bar{\ssstyle M \kern -1pt S}}}

\begin{document}
\Chapter{The Physics Program at MAMI-C}
           {The Physics Program at MAMI-C}{Patrick Achenbach}
\vspace{-5.5 cm}\includegraphics[width=6 cm]{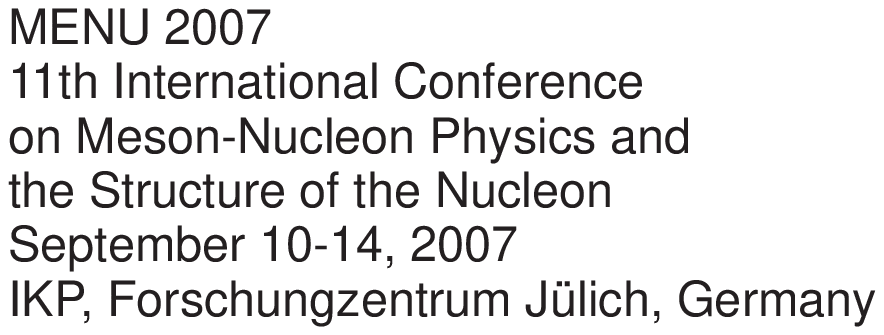}
\vspace{4 cm}

\addcontentsline{toc}{chapter}{{\it P. Achenbach}} \label{authorStart}

\begin{raggedright}
  {\it Patrick Achenbach}\index{author}{Achenbach, Patrick}\\
  Institut f\"ur Kernphysik\\
  Johannes Gutenberg-Universit\"at\\
  J.-J.-Becher-Weg 45\\
  D-55099 Mainz, Germany\\
  Email: patrick@kph.uni-mainz.de\bigskip\bigskip
\end{raggedright}

\begin{center}
  \textbf{Abstract}
\end{center}
In February 2007, the fourth stage of the Mainz Microtron, MAMI-C,
started operations with a first experiment. The new Harmonic
Double-Sided Microtron delivers an electron beam with energies up to
1.5\,GeV while preserving the excellent beam quality of the previous
stages. The experimental program at MAMI is focused on studies of the
hadron structure in the domain of non-perturbative QCD. In this paper,
a few prominent selections of the extensive physics program at MAMI-C
will be presented.

\index{subject}{Photon and electron interactions with hadrons}
\index{subject}{Protons and neutrons}
\index{subject}{Electromagnetic processes and properties}
\index{subject}{Meson production}
\index{subject}{Cyclic accelerators}

\section{The new 1.5\,GeV Harmonic Double-Sided Microtron as the
  Fourth Stage of MAMI}
\begin{figure}
  \begin{center}
    \includegraphics[width=0.6\textwidth]{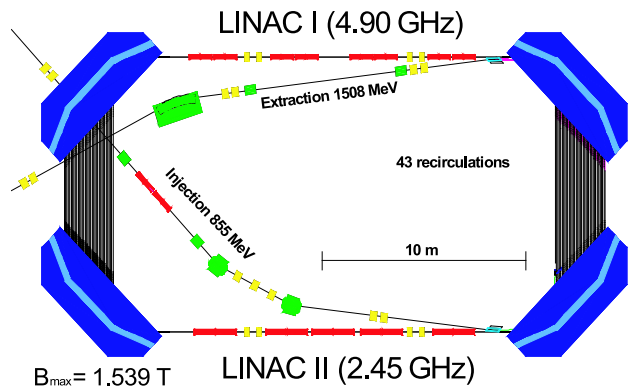}\vspace{-4mm}
    \caption{Scheme of the Harmonic Double-Sided Microtron for MAMI-
      C~\cite{Janko2006}.\vspace{-5mm}}
    \label{fig:MAMI-C}
  \end{center}
\end{figure}

The Mainz Microtron MAMI is a unique facility in Europe to study the
hadron structure with the electromagnetic probe at small momentum
transfers~\cite{SFB0810}. MAMI comprises a cascade of three race-track
microtrons (RTM), delivering since 1991 a high-quality 855\,MeV,
100\,$\mu$A cw-electron beam~\cite{Euteneuer1994}, which was energy
upgraded by a fourth stage, MAMI-C, in recent years~\cite{Janko2006}.
Realizing the fourth stage as another RTM was evidently impossible:
the two 180$^\circ$-bending magnets would have had a weight of
approx.\ 2\,$\times$\,2000\,tons for an end-point energy of
1.5\,GeV. However, for the next higher polytron configuration, the
Double-Sided Microtron (DSM), this weight is reduced by a factor of
four. By operating at 4.90\,GHz, the first harmonic of the fundamental
2.45\,GHz radio-frequency (rf), the necessary coherent energy gain per
turn for the two normal conducting linacs is reduced to get a moderate
power consumption. For achieving a simple transverse optics the strong
vertical fringe field defocusing of the four 90$^\circ$-bending
magnets is compensated by a clam-shell field geometry. In this
configuration uncritical longitudinal beam dynamics is achieved by
operating one of the two linacs at the fundamental rf. The
acceleration by two different rf in the Harmonic Double-Sided
Microtron (HDSM) is possible because in one of its linacs only every
second 4.90\,GHz-bucket is occupied~\cite{Kaiser1999,Janko2002}. A
scheme for the HDSM is given in Fig.~\ref{fig:MAMI-C}.

The commissioning of the HDSM started in autumn 2006 with the first
full turn through the HDSM. The acceleration along the full 43
recirculations was first operational at the end of 2006, and since
February 2007 MAMI-C is delivering a 1.5\,GeV high power beam of
polarized electrons to the experimental areas.

\section{Selection A: Polarized $\eta$ Electroproduction}
The electromagnetic production of $\eta$ mesons is a selective probe
to study the resonance structure of the nucleon. The polarized target
asymmetry was measured in Bonn at the PHOENICS
experiment~\cite{Bock1998}. This measurement showed a surprising
angular structure, which cannot be described by the existing
phenomenological models. A detailed model-independent
study~\cite{Tiator1999} showed that one possibility to describe these
data is to include a strong phase shift between $s$- and $d$-waves.

\begin{figure}
  \begin{center}
    \includegraphics[width=0.6\textwidth]{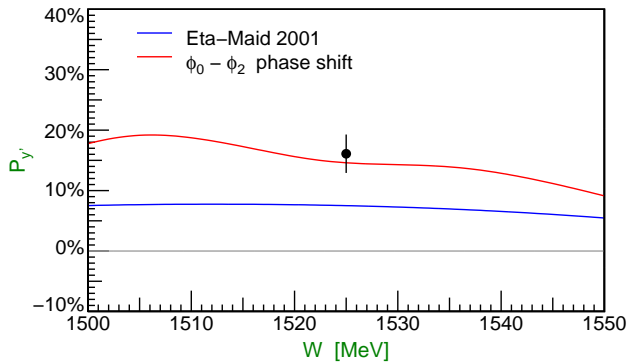}\vspace{-4mm}
    \caption{Recoil polarization observable $P_{y'}$ as a function of
      the c.m.\ energy $W$ at $\theta$ = 120$^\circ$, $Q^2$ =
      0.1\,GeV$^2/c^2$, and $\epsilon$ = 0.718 from
      Ref.~\cite{Merkel2007}. The solid line shows the prediction of
      Eta-MAID~\cite{Chiang2002}, the dashed line shows the same model
      prediction with the energy dependent phase shift of
      Ref.~\cite{Tiator1999}.\vspace{-5mm}}
    \label{fig:py}
  \end{center}
\end{figure}

A measurement of two helicity-dependent polarizations and one
helicity-independent polarization in $\eta$ electroproduction on the
proton was performed at the three spectrometer set-up of the A1
collaboration~\cite{Merkel2007}. At $Q^2 = 0.1\,\text{GeV}^2/c^2$ the
kinematics of the experiment was chosen for an invariant mass $W$ of
the $D_{13}(1520)$ resonance. Neglecting longitudinal multipoles and
their interferences the helicity-independent polarizations are
dominated by the structure functions ${R_{T}^{y'0}} \approx$
$\sin\theta\,\Im \left\{E_{0+}^* (3 \cos\theta (E_{2-}-
  3M_{2-})-2M_{1-})\right\}$ and ${^c\!R_{TT}^{y'0}} \approx$
$3\sin\theta\cos\theta \,\Im
\left\{E_{0+}^*\left(E_{2-}+M_{2-}\right)\right\}$.  Thus, the
interference with $E_{0+}$ amplifies the sensitivity to the $d$-wave
multipoles $E_{2-}$ and $M_{2-}$. In particular, $^c\!R_{TT}^{y'0}$ is
proportional to the sine of the phase difference between $E_{0+}$ and
$E_{2-} + M_{2-}$.

The measured double polarization observables $P_{x'}^h$ and
$P_{z'}^h$, dominated by $|E_{0+}|^2$, are well described by the
Eta-MAID model.  The measured single polarization observable $P_{y'}$
disagrees with the model, see Fig.~\ref{fig:py} (solid line). However,
if a strong phase change between $E_{0+}$ and $E_{2-} + M_{2-}$, as
discussed in Ref.~\cite{Tiator1999}, is applied, the data point is in
good agreement with the model. Such a strong phase change is not easy
to achieve if one assumes a standard Breit-Wigner behavior for the
$S_{11}$(1535) resonance.

\section{Selection B: Physics with the Photon Beam}
In order to deal with the MAMI-C end-point energy increase, the photon
tagging system has been extended and refurbished. The Crystal Ball
detector, a photon spectrometer consisting of 672 NaI crystals, has
been installed at the photon beam-line in recent
years~\cite{Watts2005}. It is now being used regularly with an inner
detector for tracking and the forward photon spectrometer TAPS for a
4$\pi$ angular coverage, as shown in Fig.~\ref{fig:CB}. A new data
acquisition system with high-rate performance is in operation and has
successfully taken high statistics data~\cite{Arends2006}. Further, a
new frozen spin target of liquid $^1\vec{\text H}$ and $^2\vec{\text
  H}$ is now being commissioned, with a cryostat temperature of 70\,mK
first reached in June 2007.  This target, in particular, provides a
unique opportunity to measure the partial contributions to the GDH sum
rule on a neutron target. An incomplete list of topics being addressed
by this instrumentation is given here, details can be found in
Ref.~\cite{SFB0810}:

\begin{figure}
  \begin{center}
    \includegraphics[width=0.6\textwidth]{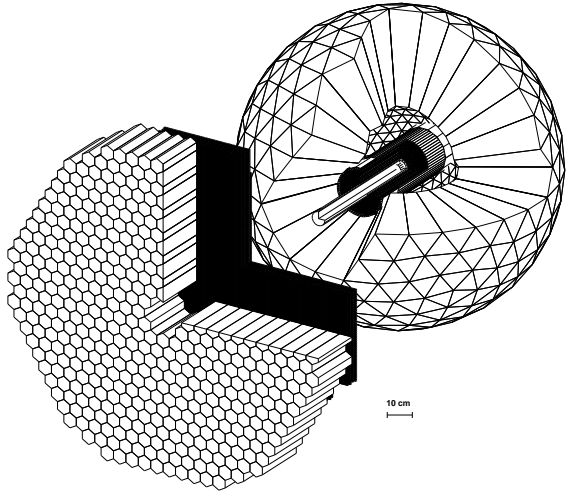}
    \vspace{-4mm}
    \caption{The Crystal Ball detector with the inner tracker and TAPS
      as forward wall~\cite{Arends2006}.\vspace{-5mm}}
    \label{fig:CB}
  \end{center}
\end{figure}

Chiral perturbation theory has been successfully applied to pion
photoproduction at threshold. Especially extensions to three flavors
urgently require precision data for kaon production processes.
Experiments will focus in particular on the following topics: tests of
chiral perturbation theory with $\gamma p \to \pi^0 p$ and $\gamma n
\to \pi^0 n$; photoinduced kaon production processes close to
thresholds; $\gamma p \to \eta' p$ at threshold; and low energy
constants of baryon chiral perturbation theory with dispersion
analysis of pion electroproduction.
    
A deeper understanding of the nature of resonances requires a reliable
determination not only of the mass spectrum but also of coupling
constants and decay vertexes. An essential prerequisite is the
exploration of polarization degrees in photoinduced meson production:
exploration of the Roper resonance $P_{11}(1440)$ in $\vec \gamma \vec
p \to \pi^0 p$; $S_{11}(1535)$ and $D_{13}(1520)$ in the target
asymmetry for the $\gamma p \to \eta p$ reaction; the role of the
$D_{15}(1675)$ and $P_{11}(1710)$ in $\eta$ photoproduction on the
neutron; magnetic moment of the $S_{11}(1535)$ in $\gamma p \to \eta p
\gamma'$; and properties of the $S_{11}(1535)$ and $D_{33}(1700)$
resonances in the $\gamma p \to \eta \pi^0 p$ reaction.

The dominant hadronic decay modes $\eta, \eta' \to 3 \pi$ only occur
due to the isospin violating quark mass difference $m_u - m_d$ or
small electromagnetic effects. A systematic study of such decays
offers an alternative way to study symmetries and symmetry breaking
patterns in strong interactions. Experiments will concentrate on the
main neutral decay channels of $\eta$ and $\eta'$ mesons: $\pi \pi$
and $\pi \eta$ interactions by a Dalitz plot analysis of the $\eta \to
3\pi^0$, $\eta' \to 3\pi^0$ and $\eta' \to \eta \pi^0 \pi^0$ decays;
cusp at the opening of the $\pi^0 \pi^0 \to \pi^+\pi^-$ threshold in
$\eta' \to \eta \pi^0 \pi^0$ and the $\pi\pi$ scattering length; and
anomalous $\eta$ decays and corrections to the Wess-Zumino-Witten
action at $O(p^6)$.

\section{Selection C: Spectrometry of Kaons}
The spectrometer facility in Mainz consists of three vertically
deflecting magnetic spectrometers freely rotable around a common
pivot~\cite{Blomqvist1998}. However, the detection of kaons under
small scattering angles is not possible with these spectrometers due
to the short lifetime of the kaons ($c\tau_K=$ 3.71\,m) and the long
flight path (close to 10\,m for spectrometer~C), and their limited
forward acceptance (5.6\,msr and 28\,msr). Therefore, reaction
products with strangeness under forward scattering angles need to be
analyzed with a shorter-orbit, large-acceptance spectrometer.

\kaos is a very compact magnetic dipole spectrometer with a large
acceptance in solid angle, $\Omega\approx$ 50\,msr, and in momentum,
$p_{max}/p_{min}\approx$ 2, making it suitable especially for the
detection of kaons. It was used (as KaoS) in heavy ion induced
experiments at the SIS facility (GSI) in the 1990s~\cite{Senger1993}.
For its operation at the spectrometer facility in Mainz a compact,
mobile, and adjustable platform on hydraulic positioning feet was
build. The platform with the spectrometer is being moved from an
installation position to a measurement position via a displacement
system of hydraulic pressure cylinders on skid-tracks. A return to its
parking position enables the complete coverage of the forward angle
region through the vertical spectrometers. This complex installation
was designed and constructed from 2003--6.

\begin{figure}
  \begin{center}
    \includegraphics[width=0.7\textwidth]{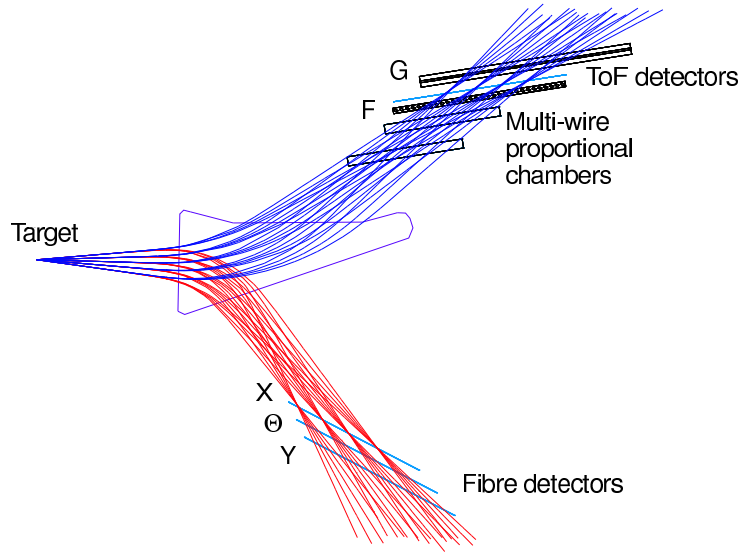}\vspace{-4mm}
    \caption{For the operation of \kaos as a double spectrometer
      trajectories of positively ($p=$ 600, 675, 750, 825\,MeV$/c$)
      and negatively charged ($p=$ 280, 310, 340, 370\,MeV$/c$)
      particles were simulated. The position of the focal planes with
      respect to the pole-face contour is indicated.\vspace{-5mm}}
  \label{fig:focalplane}
  \end{center}
\end{figure}

In November 2007 a first coincidence experiment was performed with
\kaos as hadron arm and spectrometer~B as electron arm. However, a
series of measurements, among them the detection of hypernuclei at
very forward angles, will be made possible only by the use of \kaos as
a double spectrometer. Experiments will be performed close to
production thresholds leaving very small center-of-mass energies to
the reaction particles, so that those are moving with approximate
center-of-mass velocity. The operation of \kaos as a double
spectrometer benefits of the accessibility of both pole face edges.
Simulated trajectories of positively and negatively charged particles
of four different momenta are shown in Fig.~\ref{fig:focalplane}. The
developments for the instrumentation of the electron arm are well
advanced~\cite{Achenbach2006}.

\section{Selection D: $\phi$ Meson Electroproduction}
The question of how properties of hadrons change once they are embedded
in nuclei is of fundamental interest. In particular the $\phi$ meson
provides an appealing probe for this field. Its properties in nuclei
are intimately connected to the way kaons and anti-kaons are modified
in a nuclear medium~\cite{Barz,Muehlich2003} and may provide
information on the in-medium strange-quark condensate $\langle s
\overline{s} \rangle$~\cite{Hatsuda1992}. $\phi$ mesons decaying
within nuclei can be studied via the $e^+ e^-$ as well as the $K^+
K^-$ decay channels.  Electrons practically do not interact with the
nuclear medium while the kaons interact strongly with the nucleus
themselves. Thus, studying $\phi$ decays under well defined conditions
may allow us to disentangle medium properties of the $\phi$ meson on
one hand and properties of kaons propagating within a nucleus on the
other hand.  Concerning the experimental side, the natural width of
the $\phi$ is rather narrow ($\Gamma=$ 4.26\,MeV$/c^2$) and it is not
masked by other neighboring resonances. Consequently, even the small
medium modifications of typically few percent expected for the $\phi$
mass at normal nuclear density~\cite{Muehlich2003,Klingl,Cabrera} may
be observable.

With the large solid angle \kaos the study of the coherent production
of $K^+ K^-$ pairs close to threshold is possible. The production of
$K^+K^-$ pairs will be dominated by intermediate $\phi$ meson
production~\cite{Barth,Sibirtsev}. Because of the small momentum of
the $\phi$ ($\langle p \rangle\approx$ 800\,MeV$/c$) and in view of
the small transverse momentum transfer, the two kaons are
kinematically constrained and the relative azimuth angle of the two
kaons peaks at 180$^\circ$. The experiment will take advantage of this
strong azimuthal correlation.

\begin{figure}
  \begin{center}
    \includegraphics[width=0.45\textwidth]{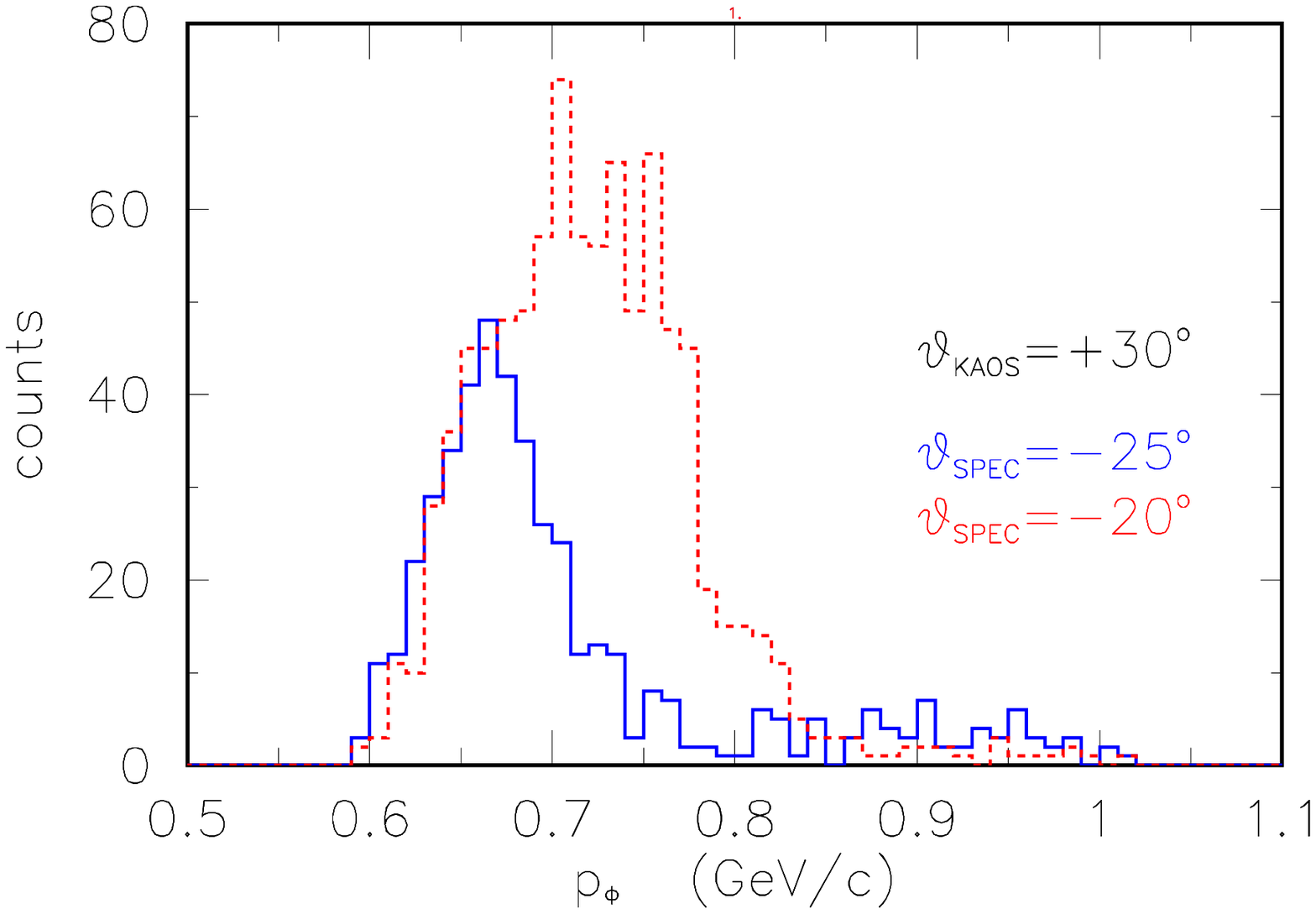}
    \hfill
    \includegraphics[width=0.45\textwidth]{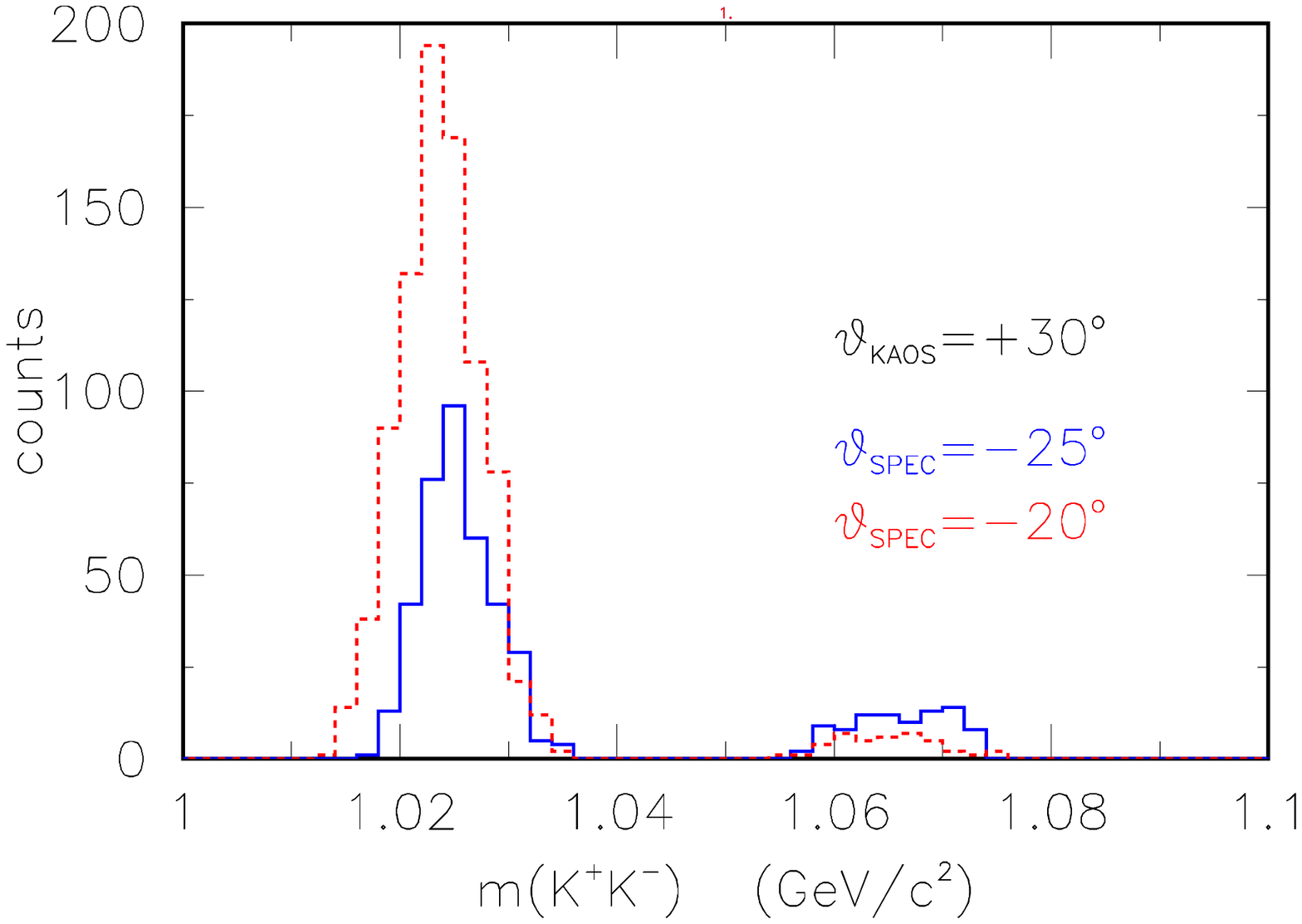}
  \end{center}\vspace{-4mm}
  \caption{Left: momentum distribution of $\phi$ mesons reconstructed
    via the $K^+ K^-$ decay channel. \kaos is positioned at an
    in-plane angle of $\vartheta=$ 30$^\circ$ and spectrometer~C
    placed at $-$20$^\circ$ (dashed histogram) or $-$25$^\circ$ (solid
    histogram). Right: reconstructed $K^+ K^-$ invariant mass.\vspace{-5mm}}
  \label{fig:kkmass}
\end{figure}

In order to explore the feasibility of a measurement of coherent
$\phi$ production at the multi-spectrometer facility of MAMI, a
schematic Monte Carlo study assuming coherent production on a
deuterium nucleus was performed.  In the left panel of
Fig.~\ref{fig:kkmass} the momentum distribution of the produced $\phi$
mesons are presented with a rather low average momentum of
approximately 0.75\,GeV$/c$. The right panel of Fig.~\ref{fig:kkmass}
shows the reconstructed $K^+ K^-$ invariant mass.  The two separated
peaks correspond to $\phi$ decays inside (right) and outside (left) of
the nucleus. The shift of the left peaks with respect to the free
$\phi$ mass of 1.019\,GeV$/c^2$ reflects a bias caused by the finite
momentum and angular acceptances for the kaons.  Comparing the dashed
and solid histograms, one recognizes that by changing the positions of
the two spectrometers the in-nucleus decays can be enhanced relative
to decays outside of the nucleus.

\section{Selection E: Two-Photon Exchange}
Recently, the discussion about second-order processes in the
electromagnetic interaction was nourished by the observation that the
ratio of proton Sachs form factors,
$R^{2}=(\mu_{p}G_{E}^{p}/G_{M}^{p})^{2}$, is different if measured by
the method of Rosenbluth separation as compared to the extraction from
the ratio of the transverse to longitudinal polarizations of the
recoiling proton~\cite{Jones2000,Gayou2002}.  A contribution from
two-photon corrections was discussed as a possible explanation for
this observation~\cite{Guichon2003}. The two-photon contributions can
be parameterized by the real part of the amplitudes
$\hat{\mathrm{G}}_{E}$, $\hat{\mathrm{G}}_{M}$ and
$\hat{\mathrm{F}}_{3}(s,Q^{2})$, where these amplitudes are
modifications of the usual Born approximation electromagnetic form
factors.

The beam normal spin asymmetry $A_{\perp}$ is an asymmetry in the
cross section for the elastic scattering of electrons with spin
parallel ($\sigma_{\uparrow}$) and spin anti-parallel
($\sigma_{\downarrow}$) to the normal polarization vector. The
evaluation of $A_{\perp}$ yields a dependence on the imaginary part of
$\hat{\mathrm{F}}_{3}(s,Q^{2})$. In contrast, the two-photon exchange
contribution to the cross section is proportional to the real part of
$\hat{\mathrm{F}}_{3}(s,Q^{2})$.  A direct ab initio calculation of
the real part of $\hat{\mathrm{F}}_{3}(s,Q^{2})$ does not seems to be
feasible. It would involve the knowledge of the off-shell form factors
of the proton in the intermediate state and also of all contributing
excitations and their off-shell transition form factors.

\begin{figure}
  \begin{center}
    \includegraphics[width=0.4\textwidth]{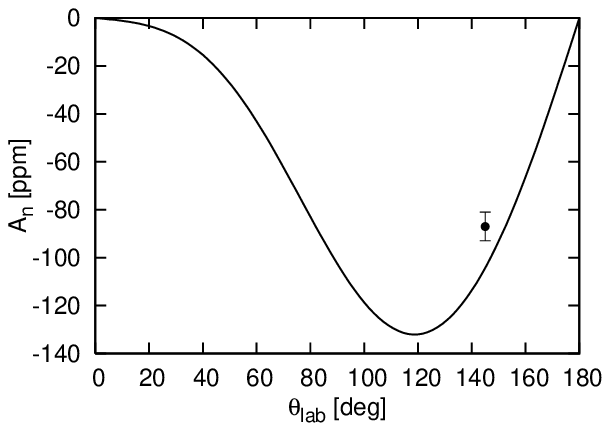}\\
    \includegraphics[width=0.4\textwidth]{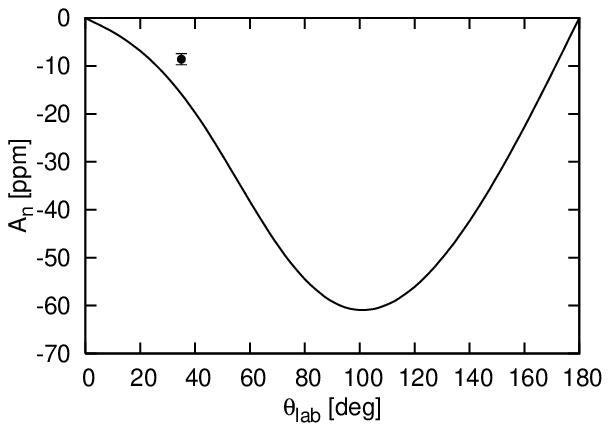}
    \includegraphics[width=0.4\textwidth]{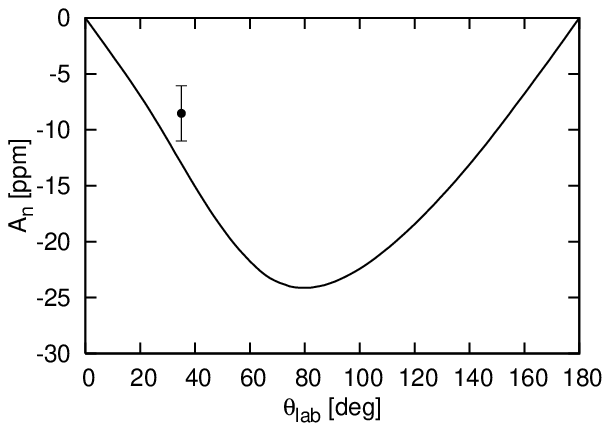}
    \vspace{-4mm}
    \caption{Results on $A_{\perp}$ at backward angles for a beam
      energy of 315\,MeV (upper panel)~\cite{SFB0810} and forward
      angles for 510\,MeV and 855\,MeV (lower panels)~\cite{Maas2005}.
      The full line is a calculation by B.~Pasquini.\vspace{-4mm}}
    \label{fig:a4}
  \end{center}
\end{figure}

The A4 collaboration has contributed to the study of the imaginary
part of the two-photon exchange amplitude by a measurement of the beam
normal spin asymmetry, $A_{\perp}$~\cite{Maas2005}. The two lower
panels of Fig.~\ref{fig:a4} show the results of measurements of
$A_{\perp}$ with the forward angle set-up ($\theta_{e}\sim
30^{\circ}-40^{\circ}$) for beam energies of 570\,MeV and 854\,MeV
together with calculations by B.~Pasquini. The upper panel contains
new data at backward angles ($\theta_{e}\sim 140^{\circ}-150^{\circ}$)
for a beam energy of 315\,MeV~\cite{SFB0810}. While the discrepancy
between measured and calculated values of $A_{\perp}$ at forward
angles could be explained by imperfect knowledge of higher mass
intermediate states, the backward angle asymmetry should be dominated
by the pure $\Delta(1232)$ excitation only. The exploration of
$A_{\perp}$ by a series of measurements at different beam energies on
proton and deuteron targets is foreseen at MAMI-C, and further studies
of the induced recoil polarization, $P_{y}$, forbidden in one-photon
exchange, are under consideration~\cite{SFB0810}.

\section{The Physics Potential of MAMI-C}
The commissioning of the new 1.5\,GeV Harmonic Double-Sided Microtron
as the fourth stage of MAMI is a great success in accelerator research
and technology.  Only a few selections from the wide range of the
physics potential at MAMI-C could be presented in this paper.  There
are further programs for testing effective field theories, {\it e.g.}\
the study of the polarizability of the nucleon and the pion, and of
generalized polarizabilities of the proton, programs for testing
models of the NN interaction, reaction mechanisms and for models of
nuclei, {\it e.g.}\ the study of few-body systems, and programs for
studying the electromagnetic and strange form factors of the nucleon.

\section*{Acknowledgments}
The physics program at MAMI is supported by the Deutsche
Forschungsgemeinschaft (DFG) via the Sonderforschungsbereich SFB443
and the European Community Research Infrastructure Activity under the
FP6 program HadronPhysics (RII3-CT-2004- 506078).




\end{document}